\documentclass[12pt]{article}
\usepackage[cp1251]{inputenc}
\usepackage[english, russian]{babel}
\usepackage{amsmath, latexsym,  amssymb, bm,  graphics, amsfonts, array, eucal}
\usepackage[dvips]{graphicx}
\makeatletter

\newcommand{\const}{\mathop{\rm const\, }}
\renewcommand{\Re}{\mathop{\rm Re}}
\renewcommand{\Im}{\mathop{\rm Im}}

\makeatother
\renewcommand{\baselinestretch}{1.1}
\newcommand{\be}{\begin{center}}
\newcommand{\ee}{\end{center}}

\begin{document}
\renewcommand{\abstractname}{\ Abstract}
\renewcommand{\refname}{\begin{center} REFERENCES\end{center}}
\newcommand{\mc}[1]{\mathcal{#1}}
\newcommand{\E}{\mc{E}}
\thispagestyle{empty}
\large

\begin{center}
\bf The kinetic theory to  problem of attenuation transversal sound waves in metal
\end{center}

\begin{center}
  \bf A. V. Latyshev\footnote{$avlatyshev@mail.ru$} and
  A. A. Yushkanov\footnote{$yushkanov@inbox.ru$}
\end{center}\medskip

\begin{center}
{\it Faculty of Physics and Mathematics,\\ Moscow State Regional
University, 105005,\\ Moscow, Radio str., 10A}
\end{center}\medskip

\begin{abstract}
Earlier electron influence on sound absorption in metal on basis of the assum\-ption
of deformation of Fermi--surfaces under the influence of a sound wave was
considered. In the present work other approach to this problem will be considered.
Our approach based on the account of dynamic (kinetic) interaction of electronic gas
with lattice fluctuations. The analysis of influence of electric field on process of
attenuation of sound fluctuations is carried out. It is shown that in the long-wave
limit this influence is essential.

{\bf Key words:} collision degenerate plasmas, Vlasov---Boltzmann equation,
atte\-nu\-ation coefficient.

{\bf PACS numbers}: 52.25.Dg Plasma kinetic equations, 52.25.-b Plasma
pro\-per\-ties, 05.30 Fk Fermion systems and electron gas.
\end{abstract}

\begin{center}
\bf  Introduction
\end{center}

Electron influence on sound absorption in metal was considered on basis of the
assumption of deformation of Fermi--surfaces under the influence of a sound wave
\cite{Ab}. However process of change of Fermi--surfaces is caused by interaction of
electronic gas with lattice and inevitably depends on characte\-ris\-tics of this
interaction. This process cannot be strictly speaking to be considered in static
approach. Dynamical and kinetical processes should be considered at the analysis of
formation of Fermi--surfaces at propagation of a sound wave in metal. In the present
work the approach to this problem will be considered, based on the account of
dynamic (kinetic) interaction of electronic gas with lattice fluctuations.

Let us consider propagation of the transversal sound in the isotropic metal. As well
as in the previous works, we will assume, that interaction of a sound wave with
electron conductivity not renders appreciable influence on speed of a sound wave.
Our purpose there will be a question consideration how electrons conductivity of
metal influence on process of attenuation of the sound wave.

The problem of attenuation of a sound wave in metal was considered in works
\cite{Landau}--\cite{Luthi}.

\begin{center}
  \bf 1. Coefficient of attenuation of a sound wave
\end{center}

The transversal sound wave creates a field of velocities in metal
$$
{\bf u}= {\bf u}_0 e^{i({\bf kr}-\omega t)},\quad {\bf ku}=0,\quad
\omega= s_{tr} k .
$$

Here $s_{tr}$ is the velocity of transversal sound oscillations, $\bf k$ is the wave
vector.

Let us choose an axis $x $ along a direction of distribution of a sound wave $ {\bf
k} =k (1,0,0) $, and an axis $y $ along a direction of velocityd $ \bf u $.
Then $ {\bf u}=u_y(0,1,0), $ where
$$
u_y=u_y(x,t)=u_0e^{i(kx-\omega t)}.
$$

The kinetic Vlasov---Boltzmann equation with integral of collisions of relaxation
type for electrons will have the following form
$$
\dfrac{\partial f}{\partial t}+\mathbf{v}\dfrac{\partial f}{\partial
\mathbf{r}}+
e\mathbf{E}\,
\dfrac{\partial f}{\partial\mathbf{p}}=\nu(f_{eq}-f),
\eqno{(1.1)}
$$
where $f_{eq}$ is the equilibrium Fermi--distribution in degenerate plasmas.

Equilibrium distribution corresponds to a condition of electronic gas which has zero
velocity comparatively of lattice. We will consider a case of isotropic metal with
undisturbed spherical Fermi--surface. The transversal sound wave leads to distortion
of lattice, that in turn can lead to distortion equilibrium Fermi--surface of metal.
In the given work we will not to be consider this effect. Thus, in the considered
case we have

$$
f_{eq}(\mathbf{v})=\Theta(\E_F-\E(\mathbf{v-u})).
$$

Here $\E(\mathbf{v-u})$ is the electron energy,
$$
 \E(\mathbf{v-u})=\dfrac{m(\mathbf{v}-\mathbf{u})^2}{2},
$$
$\Theta(x)$ is the unit step of Heaviside,
$$
\Theta(x)=1,\quad x>0;\qquad \Theta(x)=0,\quad x<0,
$$
quantities $e$ and $m$ are charge and mass of electron сorrespondingly.

At such statement of this problem we consider, that electrons at scat\-te\-ring to pass
in equilibrium condition with the lattice which possesses the local speed $
\mathbf{u} = \mathbf{u}(x, t) $.

At such approach to this problem there is no necessity to enter "fictitious forces"
\, \cite {Pippard}, responsible for interaction electronic gas with lattice
fluctuations.

At considered statement of the problem electric field $ \mathbf{E} $ has only one
nonzero to a component $ {\bf E} =E_y (0,1,0) $, where
$$
 E_y=E_y(x,t)=E_0e^{i (kx-\omega t)}.
$$

Because of an electroneutrality the volume density of a charge of ions (lattice)
equals $-enu $, where $n =\const $ is the electron concentration, the current
density is equal $-en$. Then the equation on the field looks like

$$
\dfrac{\partial^2 E_y(x,t)}{\partial x^2}=
-\dfrac{4\pi i \omega}{c^2}\,\Big[j_y(x,t)-enu_y(x,t)\Big].
\eqno{(1.2)}
$$\medskip

Here $j_y(x,t)$ is the electron current density along axis $y$,
$$
j_y=e\int v_yf\dfrac{2d^3p}{(2\pi\hbar)^3}.
$$

In linear approach the distribution electron function  in the field sound wave we
will search in the form
$$
f=f_0-\dfrac{\partial f_0}{\partial \E}\psi,\qquad
\E=\E(\mathbf{v})=\dfrac{m\mathbf{v}^2}{2}.
\eqno{(1.3)}
$$

Here $f_0$ is the absolute Fermian,
$$
f_0=f_0(\E)=\Theta(\E_F-\E(\mathbf{v})).
$$

The energy flux (stream), a transferable by sound wave is equal
$$
I=\dfrac{\rho u_0^2 s_{tr}}{2},
$$
where  $\rho$ is the density of metal.

Because of attenuation of a sound wave
$$
I=I_0e^{-\Gamma_e x},
$$
where $ \Gamma $ is the coefficient of attenuation.

The coefficient of attenuation $ \Gamma_e $ is defined by following expression
$$
\Gamma_e=\dfrac{Q_e}{I}.
\eqno{(1.4)}
$$

Here $Q_e $ is the density of energy dissipation  of a sound wave.

The dissipation is caused by nonlinearity of fluctuations of a lattice and
interaction of sound fluctuations of electronic component and
field generated by a electromagnetic  wave.
The quantity $Q_e $ is calculated as follows

$$
Q_e=-F_yu_y.
\eqno{(1.5)}
$$

Here force $F_y$ is caused by two different factors.

First, on the charged ions operate the electric force induced by
electric currents $\sigma E $. Here $ \sigma $ is the volume
density of a charge of a lattice. Because of a metal electroneutrality it is had
$ \sigma =-ne $.

On the other hand on a lattice operates force of a friction from the party
of electronic gas $\bf F_e$. At each scattering of electrons, having velocity
$ \bf v$, on the average the momentum $m {\bf (u-v)} $ is  transferred to a lattice.
Considering, that force $\bf F_e$ has only one the nonzero component,
directed along the axis lengthways $y $, we have
$$
F_e=-\nu m\int(u_y-v_y)f\dfrac{2d^3{p}}{(2\pi \hbar)^3}=
-\nu u_y m\int f\dfrac{2d^3{p}}{(2\pi \hbar)^3}+$$$$+
\nu m\int v_yf\dfrac{2d^3{p}}{(2\pi \hbar)^3}=-
\nu u_y m n+\nu m n\bar v_y.
$$ \medskip

Here it has been considered, that the numerical density (concentration) of electrons is equal
$$
n=\int f d\Omega_F=\int f\dfrac{2d^3p}{(2\pi\hbar)^3},
$$
also the average quantity of $y$--components of electron velosity along an axis
$y $ is entered,
$$
\bar v_y=\dfrac{1}{n}\int v_yf\dfrac{2d^3p}{(2\pi\hbar)^3}.
$$

Thus, force of the friction is equal
$$
F_y=-\nu nmu_y-enE+\nu m n\bar v_y.
$$

Therefore the density of energy dissipation of a sound wave is the following
$$
Q_e=[enE +\nu m n (u_y-\bar v_y)]u_y.
$$ \medskip

It is obvious, that the velocity  of of electron current and average
quantity $y$--component of electrov velocity are connected among themselves
by equa\-li\-ties $j_y=en\bar v_y $ and $ \bar v_y = (1/en) j_y $.

Let us consider further, that

$$
j_y=j_y^\circ e^{i(kx-\omega t)},\qquad \bar v_y=\bar v_y^\circ e^{i(kx-\omega t)}.
$$

Let us make averaging on time (see, for example, \cite{Landau}) equalities (1.5).
Passing to the real variables, we receive

$$
Q_e=-\dfrac{1}{2}\Re(F_yu_y^*)=\dfrac{1}{2}
\Re\left\{u^*_y\left[enE+\nu m n(u_y-\bar v_y)\right]\right\}=
$$

$$
=u_0\dfrac{1}{2}
\left\{en\Re E_0+\nu m n(u_0-\Re \bar v_y^\circ)\right\}=
$$

$$
=\dfrac{u_0}{2}\left[en\Re E_0+\nu m n(u_0-\Re\bar v_y^\circ)\right].
\eqno{(1.6)}
$$\medskip

Here the top index "asterisk" \, means complex conjugation.

According to (1.4) and (1.6) coefficient of attenuation of a sound wave it is equal

$$
\Gamma_e=\dfrac{en\Re E_0+\nu m n(u_0-\Re \bar v_y^\circ)}{\rho u_0s_{tr}}.
\eqno{(1.7)}
$$\medskip

\begin{center}
  \bf 2. The solution of system of the equations and electric current density
\end{center}

For the solution of the equation (1.1) we search according to (1.3).
Transform (1.3), we receive, that
in the case of degenerate plasmas in metal distribution function in linear approach
it is searched in the form

$$
f=f_0(\E)+\delta(\E_F-\E)\psi,
\eqno{(2.1)}
$$
where $\psi=\psi(x,\mathbf{v},t)$ is the new unknown function.

After linearization of locally equilibrium distribution function on
quantity of speed $ {\bf u} $ we receive

$$
f_{eq}=f_0(\E)+mv_y u_y \delta(\E_F-\E).
\eqno{(2.2)}
$$

By means of (2.1) and (2.2) the equation (1.1) it will be transformed
to the following equation

$$
\dfrac{\partial \psi}{\partial t}+v_x\dfrac{\partial \psi}{\partial x}+\nu \psi=
eE_yv_y+\nu mv_yu_y.
\eqno{(2.3)}
$$

In the right part of the equation (1.2) there is a density of  current,
which taking into account equality (2.1) it is equal

$$
j_y=e\int v_yf\dfrac{2d^3p}{(2\pi \hbar)^3}=e\int v_y\psi
\delta(\E_F-\E)\dfrac{2d^3p}{(2\pi \hbar)^3}.
\eqno{(2.4)}
$$

From the equation (2.3) we find function
$$
\psi=v_y\dfrac{eE_0+\nu m u_0}{\nu-i \omega+ikv_x}e^{i(kx-\omega t)}.
\eqno{(2.5)}
$$

Let us substitute expression (2.5) in the previous expression (2.4) for electric
current density.
Then for current density it is received following expression
$$
j_y=\dfrac{2e m^3(eE_0+\nu m u_0)e^{i(kx-\omega t)} }{(2\pi\hbar)^3}\int
\dfrac{v_y^2\delta(\E_F-\E)}{\nu-i\omega+ikv_x}d^3v.
\eqno{(2.6)}
$$

Integral from (2.6) is equal
$$
\int\dfrac{v_y^2\delta(\E_F-\E)}{\nu-i\omega+ikv_x}d^3v=
\dfrac{2i\pi v_F^2}{mk_Fq^3}\varphi(q,y).
$$

Here
$$
q=\dfrac{k}{k_F}, \qquad \Omega=\dfrac{\omega}{k_Fv_F},\qquad
y=\dfrac{\nu}{k_Fv_F},\quad \dfrac{q}{y}=\dfrac{k}{\nu v_F},
$$ \medskip
$$
\varphi(q,\Omega, y)=
q(\Omega+iy)-\dfrac{1}{2}[q^2-(\Omega+iy)^2]\ln\dfrac{\Omega+iy-q}{\Omega+iy+q}.
$$\medskip

The density of longitudinal current now equals
$$
j_y=i\dfrac{2ep_F(eE_0+\nu mu_0)e^{i(kx-\omega t)}}{(2\pi\hbar)^2 q^3}\varphi(q,\Omega,y).
\eqno{(2.7)}
$$

Let us consider that fact, that velocity  of the sound is much less than
electron velocity on Fermi's surfaces:
$s _ {tr} \ll v_F $.
Then the quantity  $ \Omega $ is small parameter
$$
\Omega=\dfrac{\omega}{k_Fv_F}=\dfrac{ks_{tr}}{k_Fv_F}=q\varepsilon_1,
$$
where $\varepsilon_1$ is so small parameter,
$$
\varepsilon_1=\dfrac{s_{tr}}{v_F}\ll 1.
$$

Taking into account this fact the expression for
$ \varphi (q, \Omega, y) $ becomes simpler and looks like
$$
\varphi(q,y)=qyi-\dfrac{1}{2}(y^2+q^2)\ln\dfrac{iy-q}{iy+q}=
$$
$$
=qyi-\dfrac{1}{2}(y^2+q^2)\ln\dfrac{y+iq}{y-iq}=
qyi-\dfrac{i}{2}(y^2+q^2)\arg\dfrac{y+iq}{y-iq}=
$$
$$
=i[qy-(q^2+y^2)\arctg\dfrac{q}{y}],
$$
or
$$
\varphi(q,y)=i\varphi_0(q,y),
$$
where
$$
\varphi_0(q,y)=qy-(q^2+y^2)\arctg\dfrac{q}{y}.
$$

Now expression for density of an electric current (2.7) becomes simpler, and its amplitude
it is equal
$$
j_y^\circ=-\dfrac{2ep_F(eE_0+\nu mu_0)}{(2\pi\hbar)^2 q^3}\varphi_0(q,y).
\eqno{(2.8)}
$$\medskip

\begin{center}
\bf  3. The electric field
\end{center}

Let us return to the equation (1.2) on electric field.

Let us substitute in this equation expression of electric field, density of
current and field of electron velocity. It is as a result received the following equation
$$
-k^2E_0+\dfrac{4\pi i \omega}{c^2}j_y^\circ=\dfrac{4\pi i \omega e nu_0}{c^2},
$$
or, by mean (2.8),
$$
-E_0\Big(k^2+\dfrac{8\pi i\omega e^2p_F}{c^2(2\pi\hbar)^2q^3}\varphi_0(q,y)
\Big)=$$$$=
\dfrac{4\pi i\omega e nu_0}{c^2}\Big(1+\dfrac{2\nu mp_F}{(2\pi\hbar)^2nq^3}\varphi_0(q,y)
\Big).
\eqno{(3.1)}
$$

We transform the equation  (3.1) to the form
$$
-E_0k_F^2\Big(q^2+\dfrac{3i\Omega_p^2\varepsilon_0^2\varepsilon_1}{2q^2}
\varphi_0(q,y)\Big)=\hspace{3cm}$$$$
\hspace{3cm}=
\dfrac{4\pi i\omega e nu_0}{c^2}\Big(1+\dfrac{3y}{2q^3}\varphi_0(q,y)
\Big),
\eqno{(3.2)}
$$
where
$$
\varepsilon_0=\dfrac{v_F}{c},\quad \Omega_p=\dfrac{\omega_p}{k_Fv_F},\quad
\omega_p=\sqrt{\dfrac{4\pi e^2 n}{m}},\quad n=\dfrac{k_F^3}{3\pi^2},
$$
$\omega_p$ is the frequency of plasmas (Langmuir).

Let us result numerical calculations for three typical metals with the parametres taken from
resulted below the table.

Table

\begin{tabular}{|c|c|c|c|c|c|}
  \hline
material &sound    & plasma          & Fermi'          &wave  \\
           &  velocity,        & frequency,         &velocity            &Fermi number,\\\hline
           &  cm/sec          & 1/sec                 & cm/sec             & 1/cm\\\hline
Potassium  & $2\cdot 10^5    $ & $6.5\cdot 10^{15} $ & $8.52\cdot 10^7 $ & $7.4\cdot 10^4$ \\
Gold       & $1.74\cdot 10^5$ & $1.37\cdot 10^{16}$ & $1.4\cdot 10^8  $ & $9.2\cdot 10^7$ \\
Argentum   & $2.6\cdot 10^5$ & $0.96\cdot 10^{15}$ & $1.39\cdot 10^8 $ & $6.9\cdot 10^6$\\
  \hline
\end{tabular}       \medskip

From equation (3.2) we obtain
$$
E_0=-ieu_0\dfrac{4k_F^2s_{tr}}{3\pi c^2}
\dfrac{2q^3+3y\varphi_0(q,y)}{2q^4+3i\varepsilon\varphi_0(q,y)},
\eqno{(3.3)}
$$
or short
$$
E_0=-ieu_0\dfrac{4k_F^2s_{tr}}{3\pi c^2}\cdot D,
$$
where
$$
D=\dfrac{2q^3+3y\varphi_0(q,y)}{2q^4+3i\varepsilon\varphi_0(q,y)}.
\eqno{(3.4)}
$$

Here
$$
\varepsilon=\Omega_p^2\varepsilon_0^2\varepsilon_1=
\Big(\dfrac{\omega_p}{k_Fv_F}\Big)^2\Big(\dfrac{v_F}{c}\Big)^2\Big(\dfrac{s_{tr}}{v_F}\Big)
=\Big(\dfrac{\omega_p}{k_Fc}\Big)^2\Big(\dfrac{s_{tr}}{v_F}\Big).
$$

From resulted above the table it is easy to receive following values of small parametres

for gold
$$
\varepsilon_0=4.67\cdot 10^{-3},\qquad \varepsilon_1=1.2\cdot 10^{-3},\qquad
\varepsilon=1.25\cdot 10^{-10},
$$

for potassium
$$
\varepsilon_0=2.8\cdot 10^{-3},\qquad \varepsilon_1=2.4\cdot 10^{-2},\qquad
\varepsilon=5.3\cdot 10^{-10},
$$

for argentum
$$
\varepsilon_0=4.6\cdot 10^{-3},\qquad \varepsilon_0=1.9\cdot 10^{-3},\qquad
\varepsilon=1.7\cdot 10^{-9}.
$$

Let us take advantage of equality (3.4) and we will allocate the real and
imaginary part at fraction $D $
$$
D=\dfrac{2q^4[2q^3+3y\varphi_0(q,y)]}{4q^8+9\varepsilon^2\varphi_0^2(q,y)}-
i\dfrac{3\varepsilon\varphi_0(q,y)[2q^3+3y\varphi_0(q,y)]}
{4q^8+9\varepsilon^2\varphi_0^2(q,y)}.
\eqno{(3.5)}
$$\medskip

Therefore the quantity  $E_0$ equals
$$
E_0=-eu_0\dfrac{4k_F^2s_{tr}}{3\pi c^2}\cdot\dfrac{(2q^3+3y\varphi_0(q,y))(3\varepsilon
\varphi_0(q,y)+2q^4i)}{4q^8+9\varepsilon^2\varphi_0^2(q,y)}.
\eqno{(3.6)}
$$

From here we receive the real part of this quantity
$$
\Re E_0=-eu_0\dfrac{4k_F^2s_{tr}}{3\pi c^2}
\cdot\dfrac{(2q^3+3y\varphi_0(q,y))3\varepsilon\varphi_0(q,y)}
{4q^8+9\varepsilon^2\varphi_0^2(q,y)}.
$$\medskip

\begin{center}
  \bf 4. Coefficient of attenuation of a sound wave
\end{center}

We note that according to  (3.6) we have
$$
eE_0+\nu m u_0=\nu m u_0\Big(1+\dfrac{eE_0}{\nu m u_0}\Big)=
\nu m u_0\Big(1-i\dfrac{4e^2s_{tr}k_F^2}{3\pi \nu m c^2}D\Big)=
$$
$$
=\nu m u_0\Big(1-i\dfrac{\varepsilon}{y}D\Big).
$$

Hence, according to (2.8) for amplitude of density of an electric current it is received
$$
j_y^\circ=-\dfrac{2e \nu u_0 mp_F}{(2\pi\hbar)^2q^3}
\Big(1-i\dfrac{\varepsilon}{y}D\Big)\varphi_0(q,y).
$$

From here for average quantity of electron velocity it is had
$$
\bar v_y^\circ=-\dfrac{2\nu u_0 mp_F}{n(2\pi\hbar)^2q^3}
\Big(1-i\dfrac{\varepsilon}{y}D\Big)\varphi_0(q,y).
\eqno{(4.1)}
$$

Let us return to the formula (1.7) for coefficient of attenuation and we will present it in
the form
$$
\Gamma_e=\dfrac{Q_1+Q_2}{\rho u_0 s_{st}},
\eqno{(4.2)}
$$
where
$$
Q_1=en\Re E_0,\qquad Q_2=\nu mn(u_0-\Re \bar v_0^\circ).
$$

For expression $Q_1$ according to (3.3) -- (3.5) it is had
$$
Q_1=-u_0\nu nm\cdot\dfrac{4e^2s_{tr}k_F^2}{3\pi c^2m\nu}D_1=
-u_0\nu nm\cdot\dfrac{\varepsilon}{y}D_1,
$$
where
$$
D_1=\dfrac{(2q^3+3y\varphi_0(q))3\varepsilon\varphi_0(q,y)}
{4q^8+9\varepsilon^2\varphi_0^2(q,y)}.
$$

Let us pass to a finding $Q_2$. According to (4.1) it is found
$$
Q_2=\nu mnu_0\Big[1+\dfrac{2\nu m p_F \varphi_0(q,y)}{n(2\pi\hbar)^2q^3}\Big(1-
\dfrac{\varepsilon}{y}\Re(iD)\Big)\Big].
$$

Let us transform the previous expression to the form
$$
Q_2=\nu mnu_0\Big[1+\dfrac{3\varphi_0(q,y)y}{2q^3}\Big(1-\dfrac{\varepsilon}{y}D_1\Big)\Big].
$$\medskip

The coefficient of attenuation of a sound wave according to (4.2) is equal
$$
\Gamma_e=\dfrac{\nu mn}{\rho s_{tr}}\Big[-\dfrac{\varepsilon}{y}D_1+1+
\dfrac{3y}{2q^3}\varphi_0(q,y)\Big(1-\dfrac{\varepsilon}{y}D_1\Big)\Big]=
$$
$$
=\dfrac{\nu mn}{\rho s_{tr}}
\Big(1-\dfrac{\varepsilon}{y}D_1\Big)\Big(1+\dfrac{3y}{2q^3}\varphi_0(q,y)\Big).
\eqno{(4.3)}
$$

Let us transform the formula (4.3) to the form
$$
\Gamma_e=\dfrac{\nu mn}{\rho s_{tr}}K(q,y),
$$
where $K(q,y)$ is the dimensionless coefficient of attenuation,
$$
K(q,y)=\Big(1-\dfrac{\varepsilon}{y}D_1\Big)\Big(1+\dfrac{3y}{2q^3}\varphi_0(q,y)\Big).
\eqno{(4.4)}
$$ \medskip

Let us present this coefficient in an explicit form
$$
K(q,y)=\Bigg(1-\dfrac{3\varepsilon^2\varphi_0(q,y)}{y}\cdot
\dfrac{2q^3+3y\varphi_0(q,y)}{4q^8+9\varepsilon^2\varphi_0^2(q,y)}\Bigg)
\Big(1+\dfrac{3y}{2q^3}\varphi_0(q,y)\Big).
\eqno{(4.5)}
$$

From expression (4.5) it is visible, that the dimensionless
coefficient  of attenuation consists of two components
$$
K(q,y)=K_1(q,y)+K_2(q,y).
$$

Here $K_1(q, y) $ is the contribution to coefficient of attenuation which brings
electric field,
$$
K_1(q,y)=-\dfrac{3\varepsilon^2\varphi_0(q,y)}{y}\cdot
\dfrac{2q^3+3y\varphi_0(q,y)}{4q^8+9\varepsilon^2\varphi_0^2(q,y)},
$$
$K_2(q,y)$ is the contribution to coefficient of attenuation which brings
electronic friction with a crystal lattice,
$$
K_2(q,y)=1+\dfrac{3y}{2q^3}\varphi_0(q,y)\Bigg(1-\dfrac{3\varepsilon^2\varphi_0(q,y)}{y}\cdot
\dfrac{2q^3+3y\varphi_0(q,y)}{4q^8+9\varepsilon^2\varphi_0^2(q,y)}\Bigg).
$$

According to (4.5) dimensionless components of coefficient of attenuation are
connected by equality

$$
K_2(q,y)=1+\dfrac{3y}{2q^3}\varphi_0(q,y)\big[1+K_1(q,y)\big].
$$\medskip

Investigating behavior of dimensionless factor $K (q, y) $,
it is possible to find out, that with
the big accuracy coefficient $K (q, y) $ it is possible to replace by
coefficient
$$
K_0(q,y)=1+\dfrac{3y}{2q^3}\varphi_0(q,y).
$$

Really, we form function of relative errors
$$
Er(q,y)=\dfrac{K(q,y)-K_0(q,y)}{K(q,y)}\cdot 100\%.
$$

It is easy to see, that
$$
Er(q,y)=\dfrac{K_1(q,y)}{1+K_1(q,y)}\cdot 100\%.
$$

Let us underline, that all plots in the present work are constructed for gold.

From Figs. 1 and 2 it is visible, that the coefficient $K_0 (q, y) $
effectively approximates coefficient $K (q, y) $ at all values of wave numbers.
Comparison of plots on Fig. 1 and 2 shows, that quantity of the deviation of
coefficient  $K_0 (q, y) $ from $K (q, y) $
decreases at transition of dimensionless frequency of collisions from $10^{-4} $ to
$10^{-3}$. More below we investigate is more thin structure of dimensionless
coefficient, representing the sum of coefficients $K_1 (q, y) $ and $K_2 (q, y) $.

%\clearpage
\begin{figure}[ht]\center
\includegraphics[width=16.0cm, height=12cm]{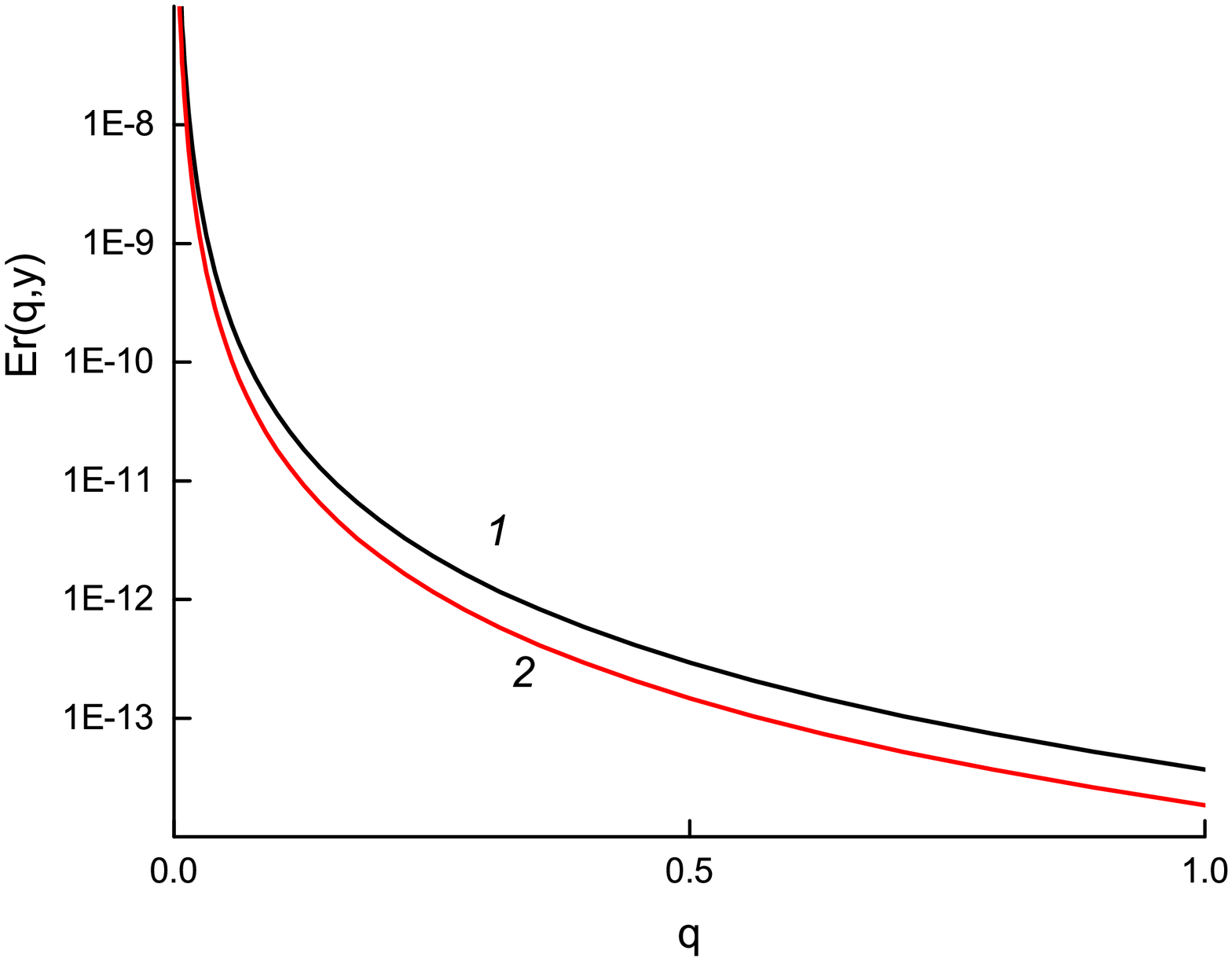}
{Fig. 1. Relative deviation of dimensionless coefficient $K_0 (q, y) $ from
coefficient $K (q, y) $. Curves of $1$ and $2$ correspond to values of frequency
$y=0.0001$ and $0.0002$.}
\end{figure}

\begin{figure}[ht]\center
\includegraphics[width=16.0cm, height=14cm]{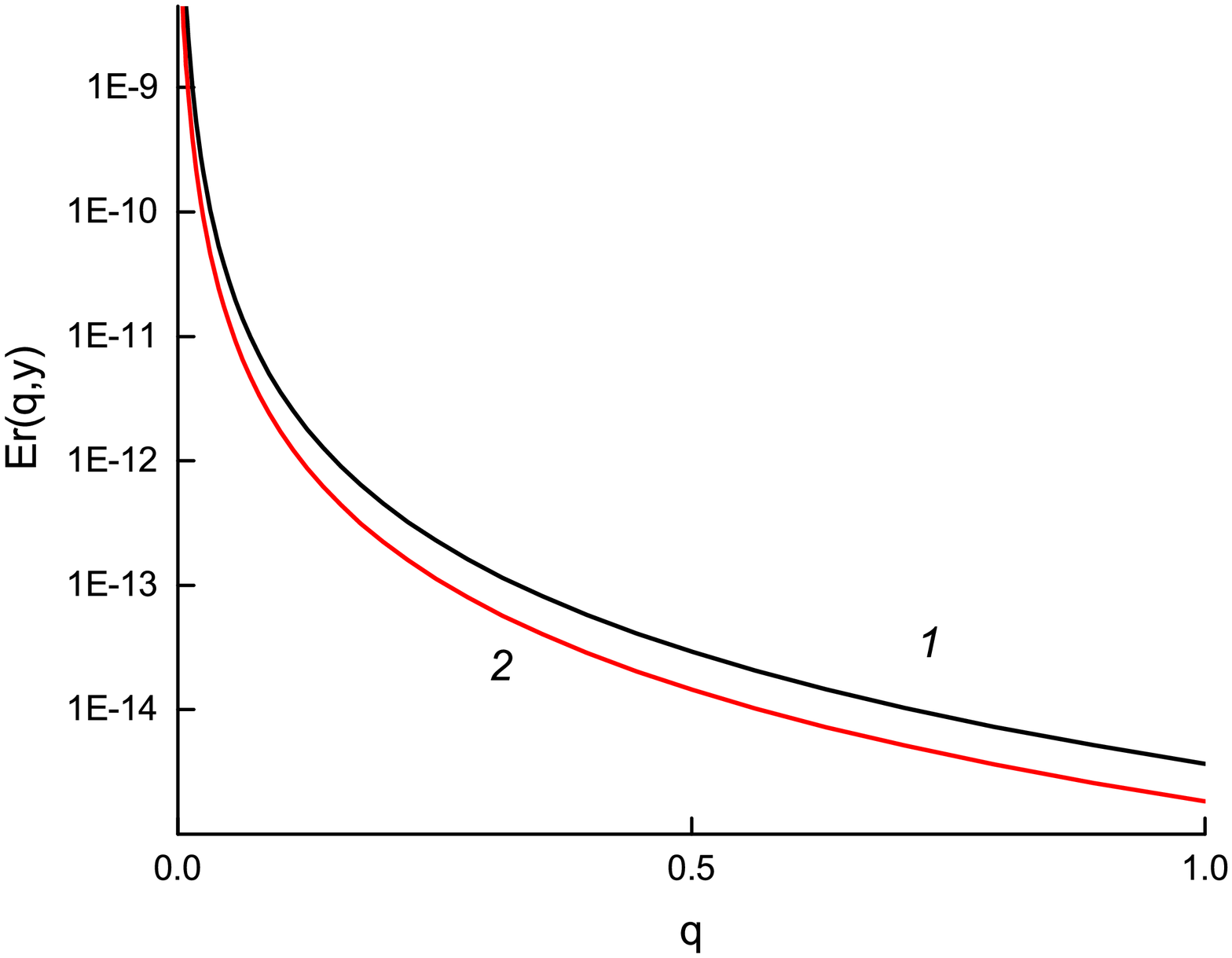}
{Fig. 2. Relative deviation of dimensionless coefficient $K_0 (q, y) $ from
coefficient $K (q, y) $. Curves of $1$ and $2$ correspond to values of frequency
$y=0.001$ and $0.002$.}
\end{figure}
\clearpage

%\newpage
%\clearpage
\begin{figure}[t]\center
\includegraphics[width=16.0cm, height=10cm]{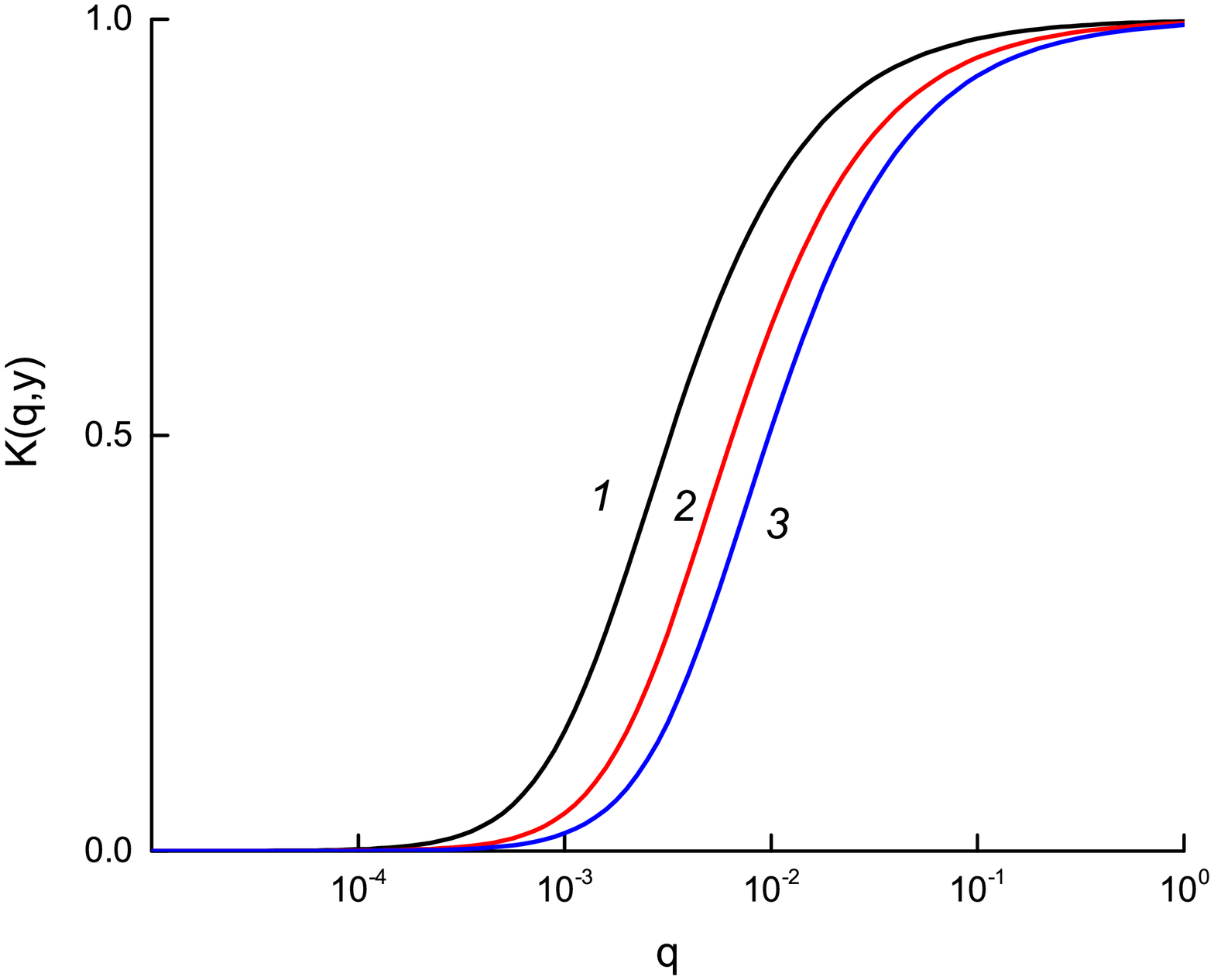}
{Fig. 3.  Coefficient of attenuation of the transversal sound wave. Curves of $1,2,3$
correspond to values dimensionless frequency of electron collisions
$y=0.001, 0.002, 0.005$.}
\includegraphics[width=16.0cm, height=10cm]{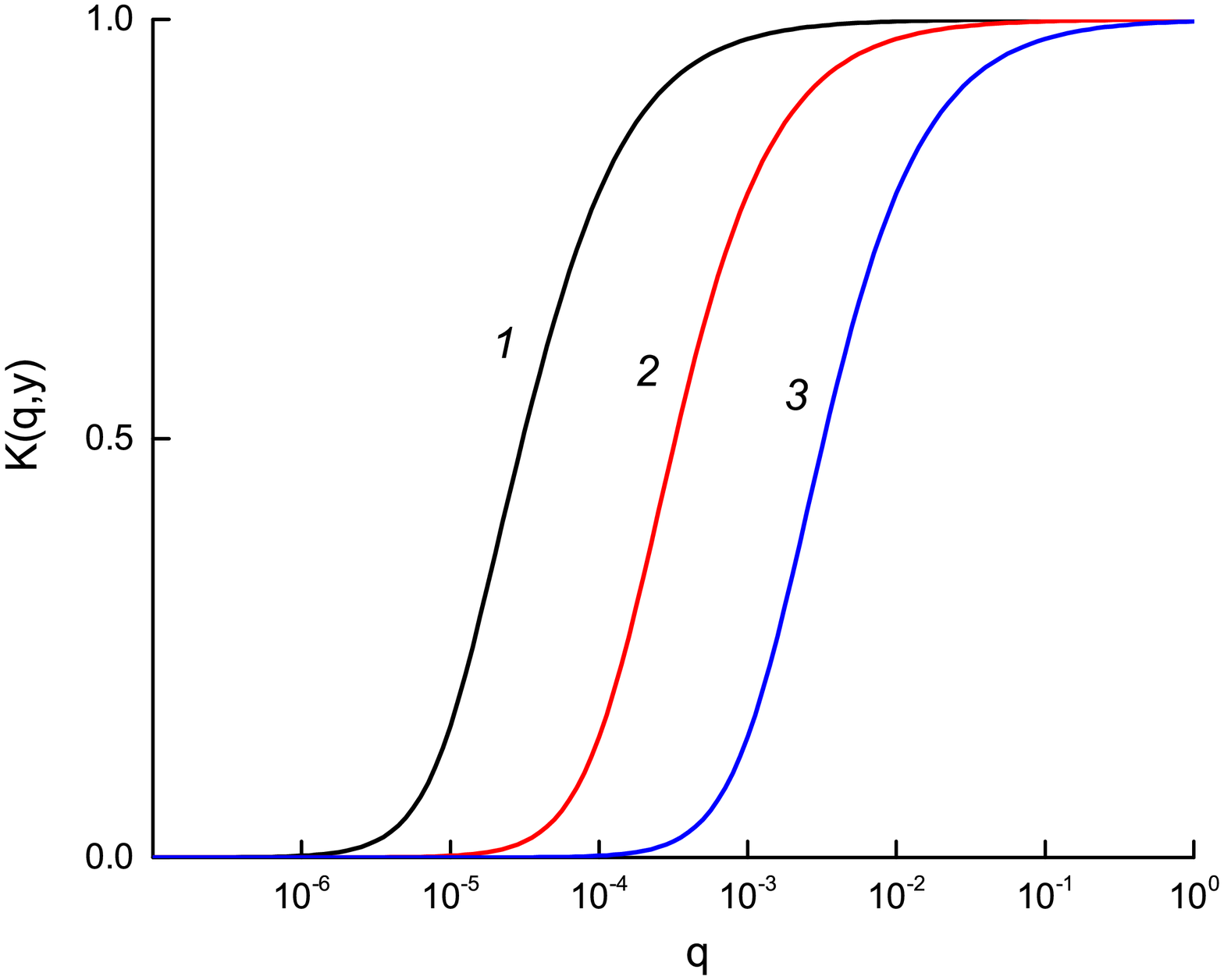}
\center{Fig. 4. Coefficient of attenuation of the transversal sound wave. Curves of $1,2,3$
correspond to values dimensionless frequency of electron collisions
$y=10^{-5}, 10^{-4}, 10^{-3}$.}
\end{figure}

\clearpage

%\newpage

\begin{figure}[t]\center
\includegraphics[width=16.0cm, height=14cm]{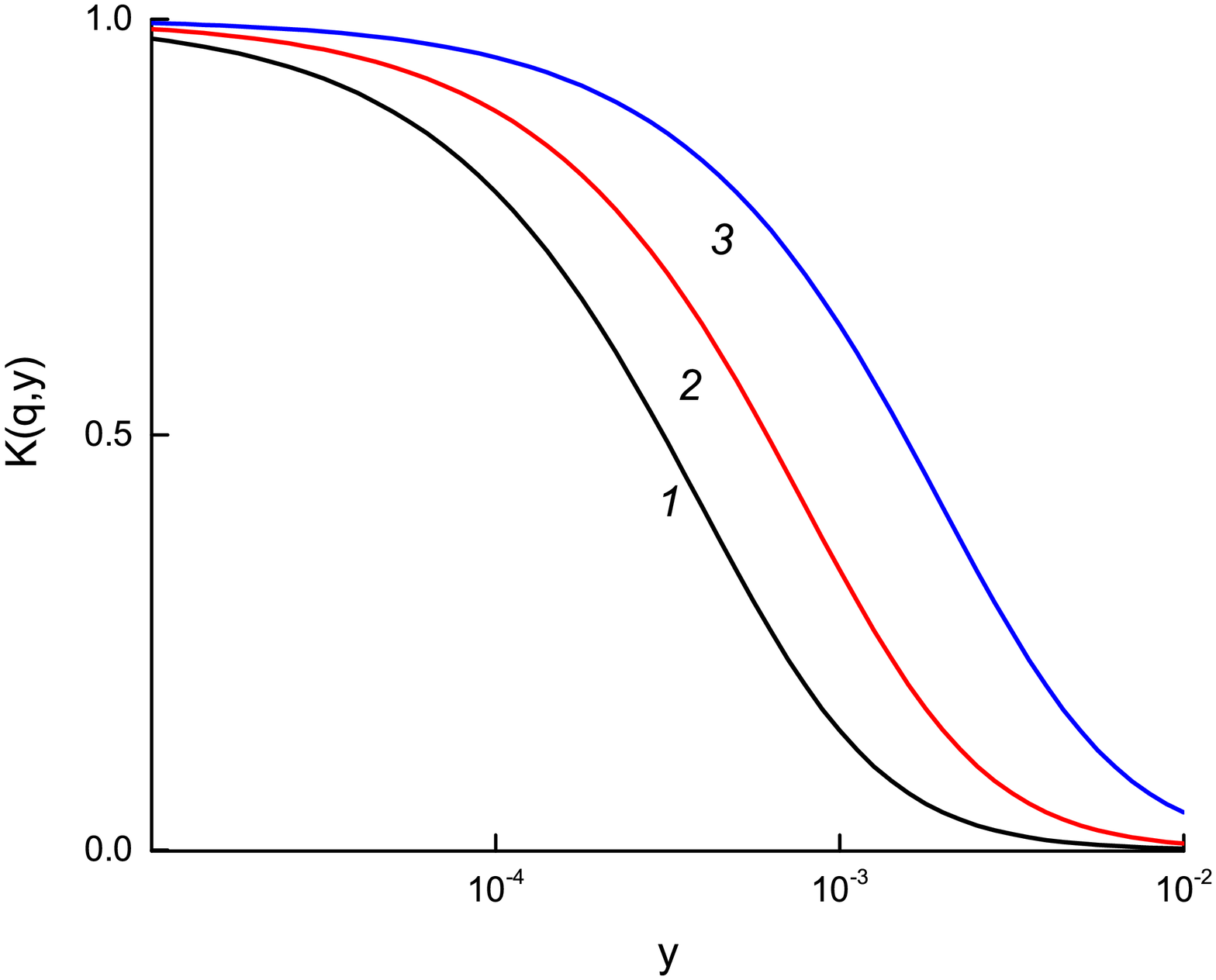}
{Fig. 5.  Coefficient of attenuation of the transversal sound wave. Curves of $1,2,3$
correspond tor values dimensionless wave number $q=0.001, 0.002, 0.005$.}
\end{figure} \bigskip

On Figs. 3 and 4 dependence of dimensionless coefficient on quantity
wave number at the fixed values of dimensionless frequency of electron collisions
is presented.
Plots on Figs. 3 and 4 show monotonous increase of coefficient of
attenuations from zero to unit.
On Fig. 5 dependence of coeffcient of attenuations from dimensionless
frequency of collisions at the fixed values
dimensionless wave number is presented.
Plots show monotonous decrease of coefficient of attenuations with growth of
frequency of collisions. \medskip \bigskip

\begin{center}
\bf 5. Coefficient of attenuations of sound wave in long--wave cases
\end{center}\medskip

Let us expand into series at small $q $ expression for $ \varphi_0 (q, y) $

$$
\varphi_0(q,y)=qy-(q^2+y^2)\Big(\dfrac{q}{y}-\dfrac{q^3}{3y^3}+\dfrac{q^5}{5y^5}-
\dfrac{q^7}{7y^7}+\cdots\Big).
$$

From here we obtain
$$
\varphi_0(q,y)=-\dfrac{2q^3}{3y}+\dfrac{2q^5}{15y^3}-\dfrac{2q^7}{35y^5}+
\dfrac{2q^9}{63y^7}\cdots=
$$
$$
=-\dfrac{2q^3}{3y}\Big[1-\dfrac{q^2}{5y^2}+\dfrac{3q^4}{35y^4}-
\dfrac{q^6}{21y^6}\Big]
$$
and
$$
2q^3+3y\varphi_0(q,y)=\dfrac{2q^5}{5y^2}-\dfrac{6q^7}{35y^4}+\dfrac{2q^9}{21y^6}\cdots=
$$
$$
=\dfrac{2q^5}{5y^2}\Big[1-\dfrac{3q^2}{7y^2}+\dfrac{5q^4}{21y^4}-\cdots\Big].
$$

Herefore the second bracet from (4.3) at small  $q$ has the following expansion
$$
1+\dfrac{3y}{2q^3}\varphi_0(q,y)=\dfrac{q^2}{5y^2}-\dfrac{3q^4}{35y^4}+
\dfrac{q^6}{21y^6}-\cdots=
$$
$$
=\dfrac{q^2}{5y^2}\Big[1-\dfrac{3q^2}{7y^2}+\dfrac{5q^4}{21y^4}-\cdots\Big].
$$

Let us return to equality (3.5) and we will consider the denominator
$
4q^8+9\varepsilon^2\varphi_0^2(q,y).
$
We have at small $q\to 0$
$$
4q^8+9\varepsilon^2\varphi_0^2(q,y)=q^6(q^2+\varepsilon^2/y^2).
$$

Let us enter critical value of dimensionless wave number $q_0$, such, that
$$
q_0=\dfrac{\varepsilon}{y}.
$$

This critical wave number breaks an interval of values of the dimensionless
wave number on two interval: the first interval $I_1 = \{0 <q <q_0 \} $
and the second interval $I_2 = \{q> q_0 \} $.

Let us estimate typical values of critical wave number for gold.
We take two values of dimensionless frequency of collisions
$y=10^{-3} $ and $y=10^{-4} $.
Corresponding values of critical wave number are equal
$$
q_0=1.25\cdot 10^{-7},\qquad \text{and}\qquad q_0=1.25\cdot 10^{-6}.
$$

From Fig. 6 it is visible, that the critical wave number is equal in accuracy
$q=q_0=1.25\cdot 10^{-6}$.

On Fig. 6 dependence of coefficients $K_1 (q, y) $ and $K_2 (q, y) $ from
wave number at the fixed value of frequency of electron collisions is presented.

From Fig. 6 it is visible, that graphics of coefficients $K_1 (q, y) $ and $K_2 (q, y) $
are crossed in the point $q_0$.
The point crossings $q_0$  of graphics is the critical value of wave number. At change
wave number from zero to $q_0$ the attenuation coefficient is defined by the electric
field (the coefficient $K_1 (q, y) $), and at $q> q_0$ the attenuation coefficient
is defined by friction of electrons on the crystal lattice (the coefficient $K_2(q, y)$).

In the first interval the dominant member of the denominator of fraction $D_1$ is
$9\varepsilon^2\varphi_0^2(q)$, in second interval the dominant member is $4q^8$.
Let us underline, that in the first interval the dominant contribution to attenuation
coefficient brings electric field (the interval of $0 <q <q_0$ on Fig. 6 see), i.e.
composed

$$
K_1(q,y)=-\dfrac{3\varepsilon^2\varphi_0(q,y)}{y}\cdot
\dfrac{2q^3+3y\varphi_0(q,y)}{4q^8+9\varepsilon^2\varphi_0^2(q,y)}.
$$\medskip

In the second interval the dominant contribution to attenuation coefficient
brings the electronic friction on the crystal lattice (the interval $q> q_0$ on
Fig. 6 see), i.e. composed
$$
K_2(q,y)=1+\dfrac{3y}{2q^3}\varphi_0(q,y)\Bigg(1-\dfrac{3\varepsilon^2\varphi_0(q,y)}{y}\cdot
\dfrac{2q^3+3y\varphi_0(q,y)}{4q^8+9\varepsilon^2\varphi_0^2(q,y)}\Bigg).
$$

Let us begin with the first interval. We will find asymptotics of attenu\-ation
coefficient in the case $0 <q\ll q_0$. We will take advantage of the previous
decomposition. At small $q\to 0$ for fraction $D_1$ we have
$$
D_1=-\dfrac{\varepsilon q^2}{5y^3}\cdot
\dfrac{\Big(1-\dfrac{q^2}{5y^2}+\dfrac{3q^4}{35y^4}-\cdots\Big)
\Big(1-\dfrac{3q^2}{7y^2}+\dfrac{5q^4}{21y^4}+\cdots\Big)}
{q^2+\dfrac{\varepsilon^2}{y^2}\Big(1-
\dfrac{2q^2}{5y^2}+\cdots\Big)}=
$$
\begin{figure}[t]\center
\includegraphics[width=16.0cm, height=12cm]{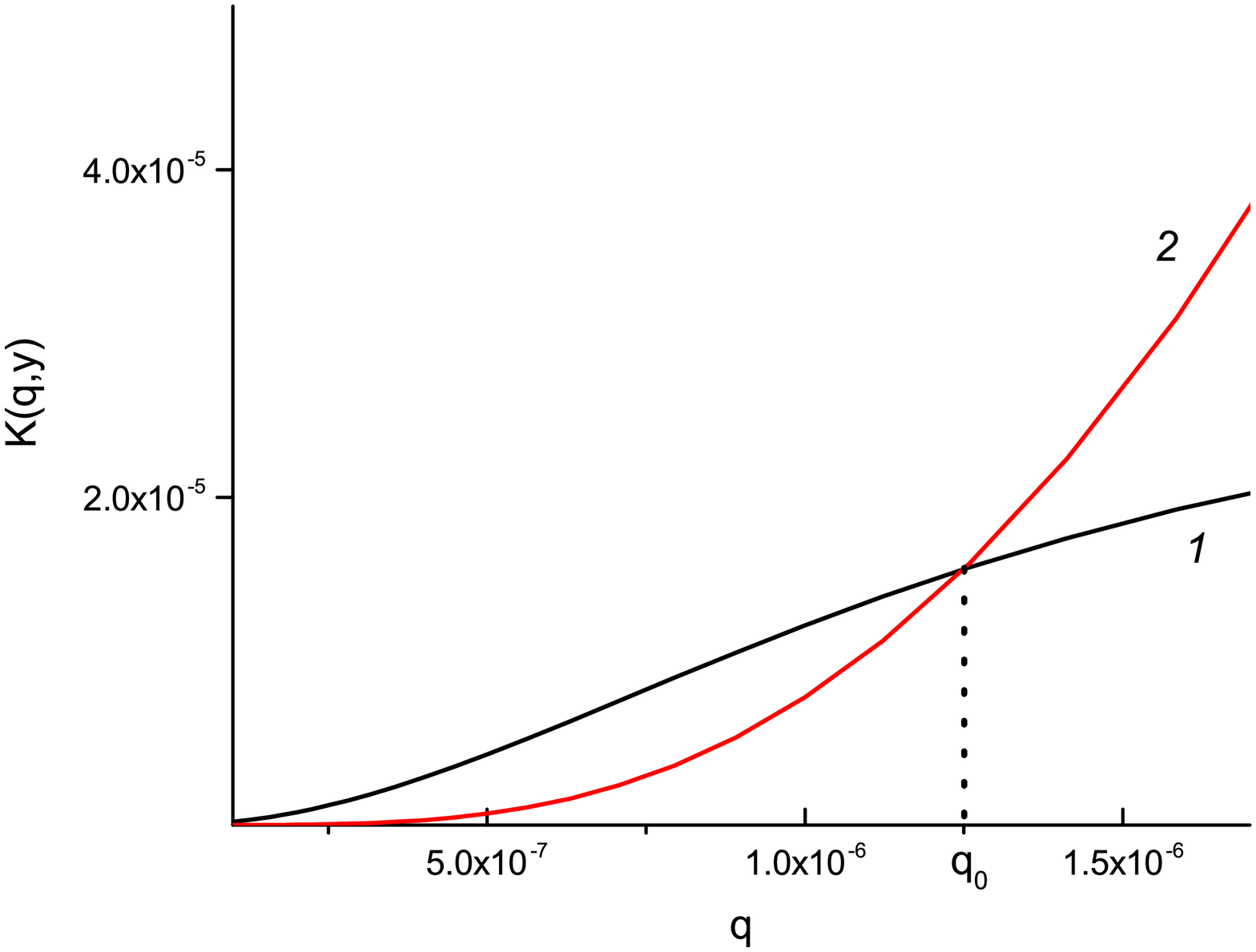}
\center{Fig. 6. The contribution to attenuation of the transversal sound wave.
Curves of $1$ and $2$  according correspond to electric field and friction of electrons
on crystal lattice, $y=10^{-4} $. At $q <q_0$ the contribution to attenuation is defined
by electric field, and at $q> q_0$ the contribution to attenuation is defined
by an electronic friction. }
\end{figure}
%\clearpage
$$
=-\dfrac{\varepsilon q^2}{5y^3}\cdot\dfrac{1-\dfrac{22q^2}{35y^2}+\cdots}
{\dfrac{\varepsilon^2}{y^2}+q^2+\cdots}=-\dfrac{\varepsilon q^2}{5y}\cdot
\dfrac{1-\dfrac{22q^2}{35y^2}+\cdots}{\varepsilon^2+q^2y^2+\cdots}.
$$

From this equality we obtain that
$$
\dfrac{\varepsilon}{y}D_1=-\dfrac{q^2}{5y^2}\cdot
\dfrac{1-\dfrac{22q^2}{35y^2}+\cdots}{1+\dfrac{q^2y^2}{\varepsilon^2}+\cdots}=
$$

$$=
-\dfrac{q^2}{5y^2}\cdot \Big(1-\dfrac{22q^2}{35y^2}+\cdots\Big)\Big(1-
\dfrac{q^2y^2}{\varepsilon^2}+\cdots\Big)=
$$
$$
=-\dfrac{q^2}{5y^2}\Big[1-\Big(\dfrac{22}{35y^2}+\dfrac{y^2}{\varepsilon^2}\Big)q^2\Big].
$$

We note that ${y^2}/{\varepsilon^2}=1/q_0^2, q_0\ll y$,
therefore in the considered interval
$$
\dfrac{22}{35y^2}+\dfrac{y^2}{\varepsilon^2}=\dfrac{22}{35y^2}+\dfrac{1}{q_0^2}=
\dfrac{1}{q_0^2}.
$$

Therefore finally we have
$$
\dfrac{\varepsilon}{y}D_1=
-\dfrac{q^2}{5y^2}\Big[1-\dfrac{q^2y^2}{\varepsilon^2}\Big]=
-\dfrac{q^2}{5y^2}\Big(1-\dfrac{q^2}{q_0^2}\Big).
$$

From the formula (4.1) for average electron velocity at small values of wave number
we receive
$$
\Re \bar v_y^\circ=-\dfrac{2\nu u_0mp_F}{n(2\pi\hbar)^2q^3}
\Big(1-\dfrac{\varepsilon}{y}D_1\Big)\varphi_0(q,y),
$$
or
$$
\Re \bar v_y^\circ=\dfrac{4\nu u_0 mp_F}{3ny(2\pi\hbar)^2}
\Big(1-\dfrac{q^2}{5y^2}+\dfrac{3q^4}{35y^4}+\cdots\Big)
\Big(1-\dfrac{\varepsilon}{y}D_1\Big).
$$

From here we find that
$$
\Re \bar v_y^\circ=u_0\Big(1-\dfrac{q^2}{5y^2}+\dfrac{3q^4}{35y^4}+\cdots\Big)
\Big(1-\dfrac{\varepsilon}{y}D_1\Big).
$$
Further we receive that
$$
\Re \bar v_y^\circ=u_0\Big(1-\dfrac{q^2}{5y^2}+\dfrac{3q^4}{35y^4}+\cdots\Big)
\Big(1+\dfrac{q^2}{5y^2}-\dfrac{q^4}{5\varepsilon^2}+\cdots\Big),
$$
from which
$$
\Re \bar v_y^\circ=u_0\Big(1-\dfrac{q^4}{5\varepsilon^2}+\cdots\Big).
$$

From here it is visible, that on absolute value in the limit at $q\to 0$ average
electron velocity tends to  velocity of the lattice
$$
\lim\limits_{q\to 0}\Re \bar v_y^\circ=u_0.
$$

Now we will find the coefficient $K (q, y) $. According to (4.4) it is had
$$
K(q,y)=\Big(1-\dfrac{q^4}{5\varepsilon^2}-\cdots\Big)
\Big(\dfrac{q^2}{5y^2}-\dfrac{3q^4}{35y^4}+\dfrac{q^6}{21y^6}-\cdots\Big)=$$$$=
\dfrac{q^2}{5y^2}-\dfrac{3q^4}{35y^4}+\cdots=
\dfrac{q^2}{5y^2}\Big[1-\dfrac{3q^2}{7y^2}+\cdots\Big].
\eqno{(5.1)}
$$

So, at small values of wave number the attenuation coefficient is equal
$$
\Gamma_e(q,y)=\dfrac{\nu nm}{\rho s_{tr}}\Big(\dfrac{q^2}{5y^2}-
\dfrac{3q^4}{35y^4}+\cdots\Big)=$$
$$
=\dfrac{\nu nm q^2}{5\rho s_{tr} y^2}\Big(1-\dfrac{3q^2}{7y^2}+
\cdots\Big),
$$
or, with dimensional parameters,
$$
\Gamma_e(k,\nu)=\dfrac{n mv_F^2k^2}{5\rho\nu s_{tr}}
\Big(1-\dfrac{3v_F^2k^2}{7\nu^2}\Big).
$$

From here in square-law approach (on $k $) $ \Gamma_e (k) = \Gamma_0k^2 \; (k\to 0) $, where
$$
\Gamma_0=\dfrac{n mv_F^2}{5\rho\nu s_{tr}}=
\dfrac{mv_F^2k_F^3}{15\pi^2\rho s_{tr}\nu}.
$$

Let us estimate one after another this quantity for gold. The mass of electron
is equal
$m=9\cdot 10^{-28} $ {\it gr}, the gold density is equal $ \rho=19.32  gr/cm^3$,
we consider, that $ \nu\sim 10^{13}\, 1/sec $,
other data we take from resulted above the table. We receive, that

$$
\Gamma_0=\dfrac{9\cdot 10^{-28}(1.4\cdot 10^8)^2(8.2\cdot 10^7)^3}
{15 \pi^2(19.3)10^{13}(1.7\cdot 10^5)}\sim 2\cdot 10^{-9}\; \dfrac{1}{cm}.
$$ \medskip

On Fig. 7 the behaviour of factors $K_1 (q, y) $ and $K_2 (q, y) $ is presented.
We will notice, that in the range of wave numbers, biger critical,
the coefficient $K_1(q, y)$ has the unique maximum, and then quickly decreases to zero.
From Fig. 8 it is visible, that the maximum quantity decreases with growth
of frequency of electron collisions.

We investigate separately behaviour of dimensionless coefficient
$K_1 (q, y) $ and $K_2 (q, y) $ at $q\to 0$.

\begin{figure}[h]\center
\includegraphics[width=16.0cm, height=14cm]{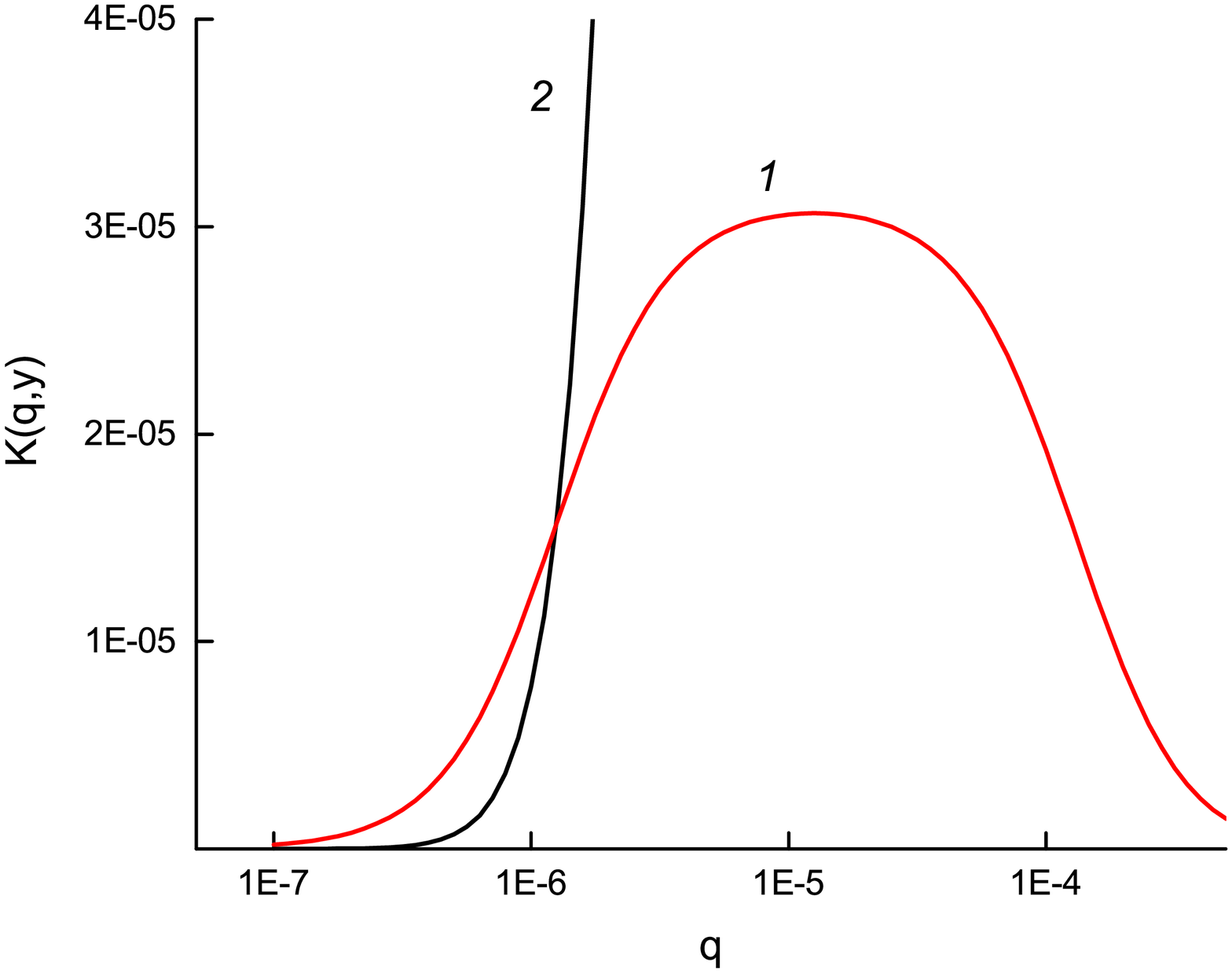}
{Fig. 7. Dependence of dimensionless coefficients of attenuation $K_1(q, y)$
(the curve 1) and $K_2(q, y)$ (the curve 2) on dimensionless frequency of collisions,
$y=0.0001$.}
\end{figure}
\clearpage

\begin{figure}[t]\center
\includegraphics[width=16.0cm, height=14cm]{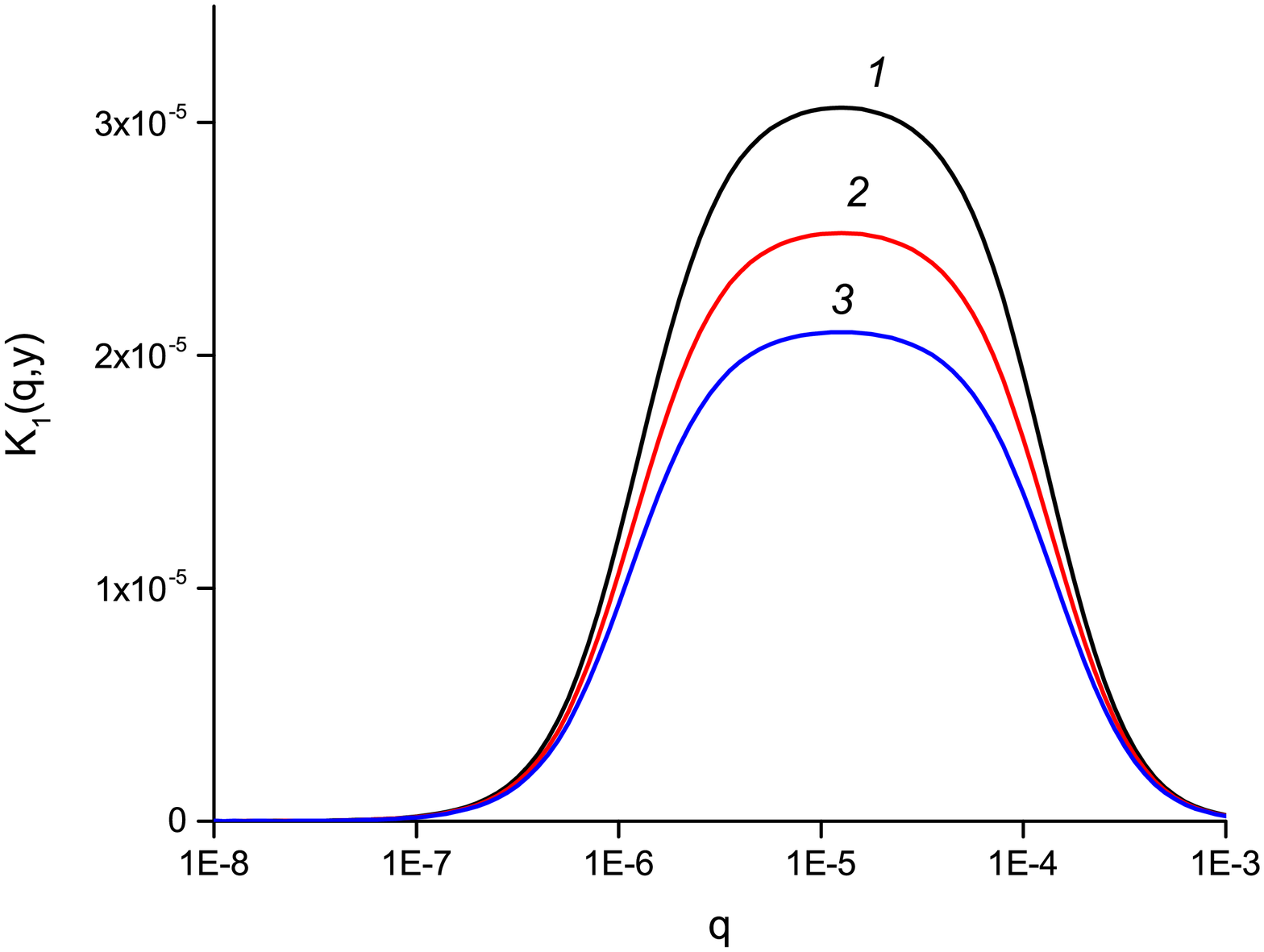}
{Fig. 8. Dependence of dimensionless coefficient of attenuation $K_1(q,y)$
on dimensionless collision frequency. Curves $1,2,3$  correspond to values of
frequency $y=0.0001, 0.000105, 0.00011$.}
\end{figure}

Let us begin with the first factor. We will present it in the form
$$
K_1(q,y)=-6\dfrac{\varepsilon^2}{y}\varphi_0(q,y)q^3\dfrac{1+\dfrac{3y}{2q^3}\varphi_0(q,y)}
{4q^8+9\varepsilon^2\varphi_0^2(q,y)}.
$$

We note that
$$
\varphi_0^2(q,y)=\dfrac{4q^6}{9y^2}\Big[1-\dfrac{2q^2}{5y^2}+
\dfrac{37q^4}{175y^4}-\cdots\Big].
$$

Not resulting calculation, we receive, that
$$
K_1(q,y)=\dfrac{q^2}{5y^2}\Big[1-\Big(\dfrac{y^2}{\varepsilon^2}+
\dfrac{3}{7y^2}\Big)q^2+\cdots\Big]=\dfrac{q^2}{5y^2}
\Big(1-\dfrac{y^2}{\varepsilon^2}q^2\Big),
$$
or,
$$
K_1(q,y)=\dfrac{q^2}{5y^2}\Big(1-\dfrac{q^2}{q_0^2}\Big),\qquad q\to 0.
$$

This expression in accuracy coincides with expression for coefficient $K(q,y)$,
a definiendum (5.1).

For  finding asymptotics of  the second coefficient we will take advantage of the formula
$$
K_2(q,y)=1+\dfrac{3y}{2q^3}\varphi_0(q,y)(1+K_1(q,y)).
$$

We obtain that

$$
K_2(q,y)=\dfrac{q^4}{5y^2}\Big(\dfrac{y^2}{\varepsilon^2}+\dfrac{1}{5y^2}\Big)=
\dfrac{q^4}{5\varepsilon^2},\qquad q\to 0.
$$

Thus, at small values of wave number the contribution to attenuation coefficient
it is defined by  presence of electric field.

Let us consider the second interval of values of dimensionless wave number $q> q_0$. This
interval we will break into three intervals: $q_0\ll q \ll y $, $y\ll q\ll 1$ and $q\sim 1$.

Let us begin with an interval $q_0\ll q\ll y $. In the considered interval

$$
K(q,y)=1+\dfrac{3y\varphi_0(q,y)}{2q^3}=
\dfrac{q^2}{5y^2}\Big(1-\dfrac{3q^2}{7y^2}+\dfrac{5q^4}{21y^4}-\cdots\Big).
$$

This formula means, that the damping coefficient is computed under the same formula,
as in the previous case.

Let us observe the third case small $q $: $y\ll q \ll 1$. In this case

$$
\arctg\dfrac{q}{y}\approx \dfrac{\pi}{2},
$$
$$
\varphi_0(q,y)=qy-\dfrac{\pi}{2}(q^2+y^2)=
q^2\Big[\dfrac{y}{q}-\Big(1+\dfrac{y^2}{q^2}\Big)\dfrac{\pi}{2}\Big]\approx
$$
$$
\approx -\dfrac{\pi}{2}q^2.
$$

Hence, the dimensionless attenuation coefficient is equal
$$
K(q,y)=\Big(1+\dfrac{3y\varphi_0}{2q^3}\Big)=1-\dfrac{3\pi}{4}\cdot\dfrac{y}{q}.
$$
The dimensional attenuation coefficient is equal
$$
\Gamma_e=\dfrac{\nu n m}{\rho s_{tr}}\Big(1-\dfrac{3\pi}{4}\cdot\dfrac{y}{q}\Big).
$$

Now we will observe the case, when $q\sim 1$. In this case
$$
\varphi_0(q,y)=q^2\Big[\dfrac{y}{q}-\dfrac{\pi}{2}\Big(1+\dfrac{y^2}{q^2}\Big)\Big]=
-q^2\Big(\dfrac{\pi}{2}-\dfrac{y}{q}\Big).
$$

The dimensionless attenuation coefficient now is equal
$$
K(q,y)=1+\dfrac{3y\varphi_0}{2q^3}=1-\dfrac{3\pi}{4}\cdot \dfrac{y}{q}+
\dfrac{3\pi}{4}\cdot \dfrac{\varepsilon^2}{yq^3}=1-\dfrac{3\pi}{4}\cdot \dfrac{y}{q}.
$$

The dimensional attenuation coefficient is computed under the same formula, as in the
previous interval.

On Fig. 9 the behaviour of function $K (q, y)/q^2$ depending on the wave number
at various values of dimensionless electron collision frequencies is presented.

Graphics show dependence of coefficient $ \Gamma_0$ from magnitude of a collision frequency
of electrons. For comparison on Fig. 10 dependence of coefficient $K_1 (q, y) $ is shown
at $q\to 0$ at various values of  collision frequency of electrons.
Let us note, that magnitudes $K (q, y)/q^2$ and $K_1 (q, y) $ decrease with
growth of collision frequency.

%\newpage
\begin{figure}[t]\center
\includegraphics[width=16.0cm, height=14cm]{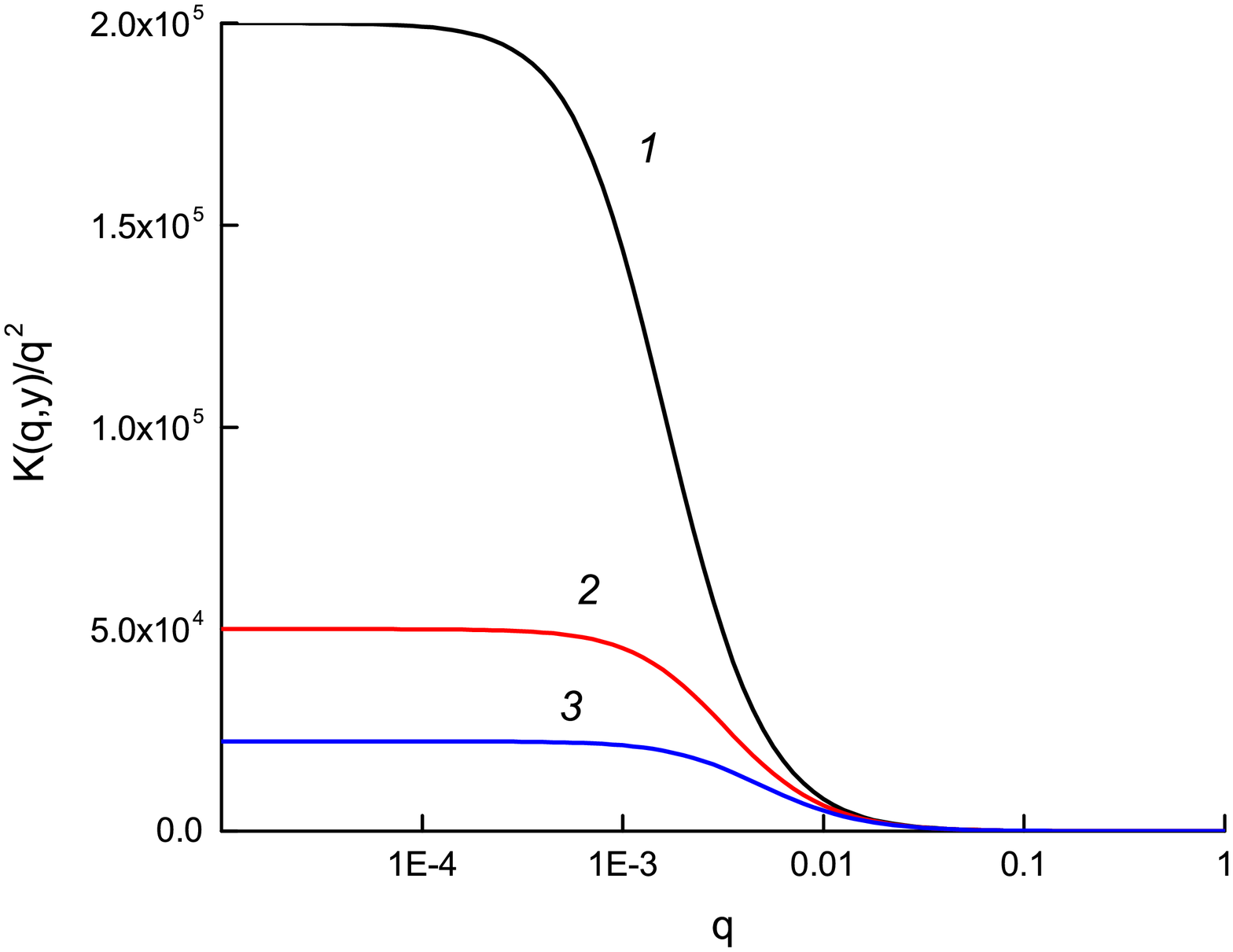}
{Fig. 9. The behaviour of quantity $K(q,y)/q^2$. Curves $1,2,3$ correspond to
values of dimensionless of electron frequency of collisions $y=0.001, 0.002, 0.003$.}
\end{figure}

%\clearpage

\begin{figure}[t]\center
\includegraphics[width=16.0cm, height=14cm]{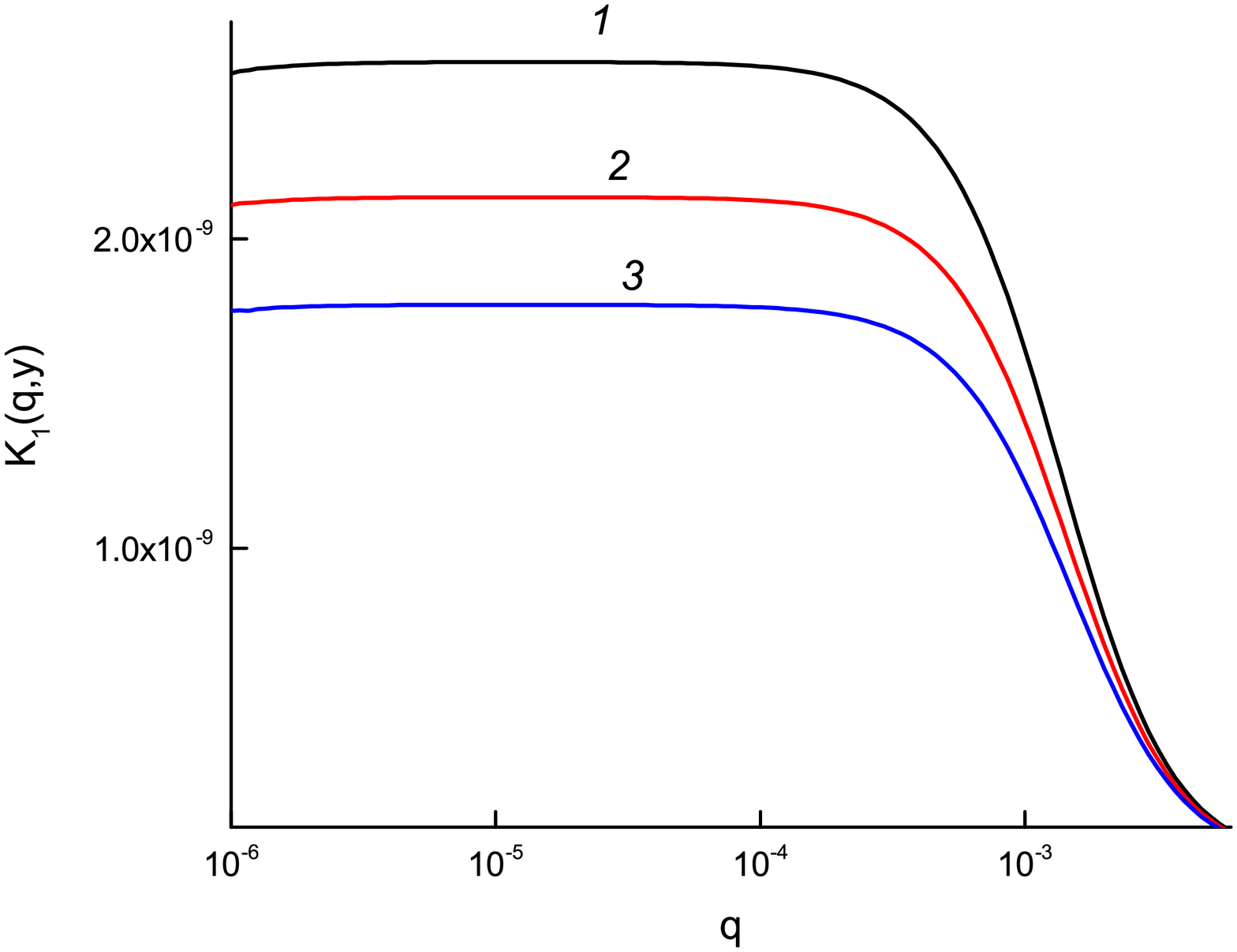}
{Fig. 10. The dependence of dimensionless coefficient of attenuation $K_1(q,y)$
on dimensionless frequancy of collisions.
Curves $1,2,3$  correspond to values of frequancy $y=0.00105, 0.00110, 0.00115$.}
\end{figure}
\clearpage

\begin{center}
  \bf 6. Comparision with Pippard' results
\end{center}

Let us result of damping coefficient discovered by Pippard \cite{Pippard}
$$
\alpha_T=\dfrac{nm}{\rho s_{tr}\tau}\Bigg[\dfrac{2}{3}\dfrac{a^3}{(1+a^2)\arctg a-a}-1\Bigg].
$$

Here
$$
a=kl=k\tau v_F=\dfrac{q}{y}.
$$

Let us this result to our denotations
$$
\alpha_T=\dfrac{\nu nm}{\rho s_{tr}}
\Bigg[\dfrac{2q^3}{3y(y^2+q^2)\arctg\dfrac{q}{y} -qy^2}-1\Bigg]=
$$
$$
=\dfrac{\nu nm}{\rho s_{tr}}\Big[-\dfrac{2q^3}{3y\varphi_0(q,y)}-1\Big].
$$

\begin{figure}[h]\center
\includegraphics[width=16.0cm, height=10cm]{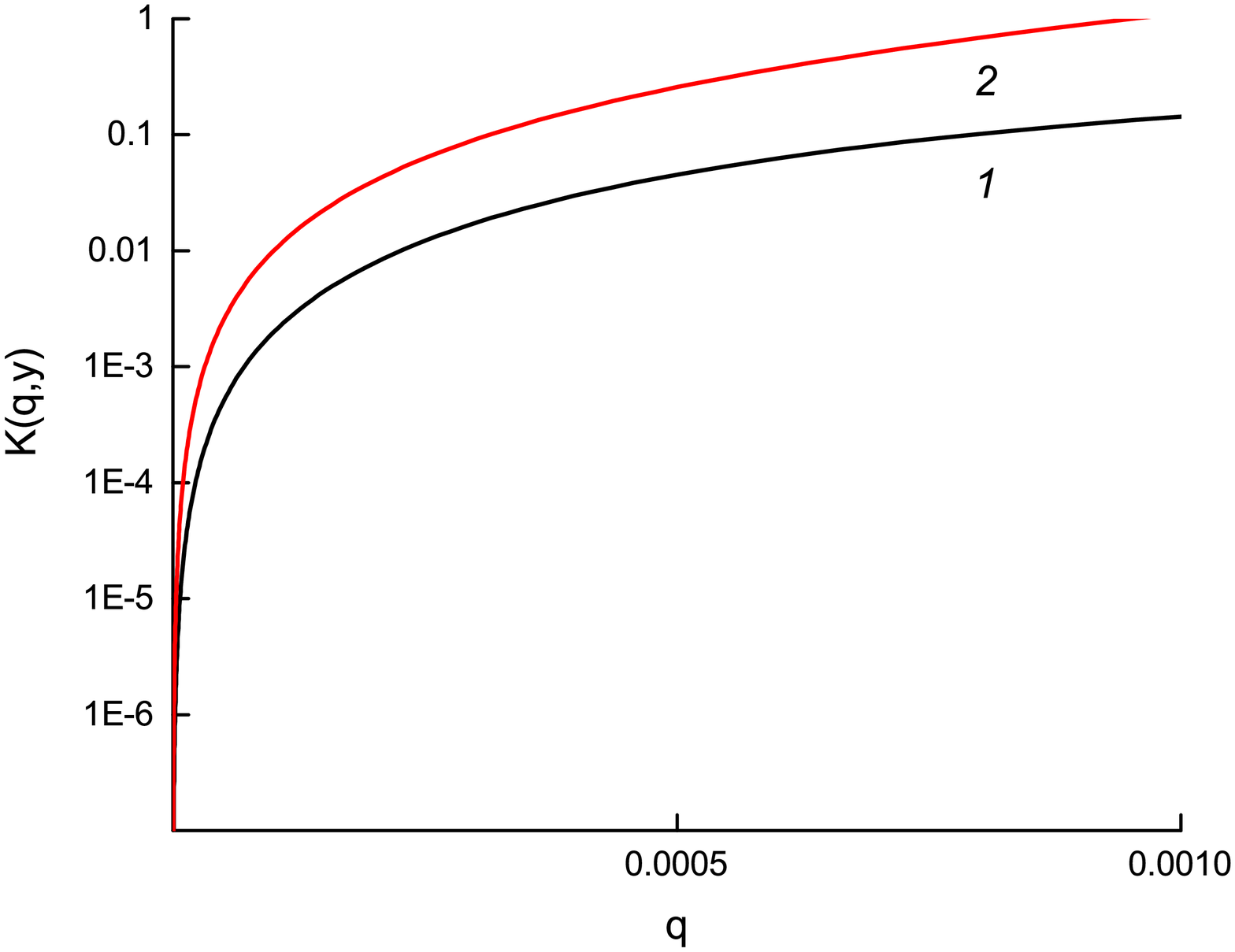}
{Fig. 11. Comparison of dimensionless coefficients of attenuation at small values
dimensionless wave number. The curve 1 corresponds to coefficient
$K (q, y) $, obtained in the present operation, the curve 2 corresponds to coefficient,
discovered by Pippard' (see \cite{Pippard}), $y=0.001$.}
\end{figure}
%\clearpage

We investigate the case of small values of the wave number. We have
$$
\alpha_T=\dfrac{\nu n m}{\rho s_{tr}}\Bigg[\dfrac{1}{1-\dfrac{q^2}{5y^2}+
\dfrac{3q^4}{35y^4}}-1\Bigg]=
$$
$$
=\dfrac{\nu n m}{\rho s_{tr}}\cdot\Bigg[\dfrac{\dfrac{q^2}{y^2}-\dfrac{3q^4}{35y^4}+
\dfrac{q^6}{21y^6}+\cdots}{1-\dfrac{q^2}{5y^2}+\dfrac{3q^4}{35y^4}-\dfrac{q^6}{21y^6}}+
\cdots\Bigg].
$$

Therefore from here we obtain
$$
\alpha_T=\dfrac{\nu n mq^2}{\rho s_{tr}y^2}\Big(1+\dfrac{4q^2}{35y^2}+\cdots\Big).
$$
This effect is comparable with our effect
$$
\Gamma_e(q,y)=\dfrac{\nu nm q^2}{5\rho s_{tr} y^2}\Big(1-\dfrac{3q^2}{7y^2}+\cdots\Big),
$$

Both coefficients diminish at $q\to 0$ as $q^2$, but with various coefficients
(Fig. 11) see.

\begin{center}
  \bf 7. Conclusion
\end{center}

In summary we will observe the case when the electric field misses. Then
$$
Q_1=0,\qquad \bar v_y^\circ=-\dfrac{2p_F\nu m u_0}{(2\pi\hbar)^2q^3}\varphi_0(q,y),
$$
$$
\Gamma_e=\dfrac{\nu n m}{\rho s_{tr}}\Big(1-\dfrac{1}{u_0}\Re \bar v_y^\circ\Big).
$$

Last expression will easily be converted to the form
$$
\Gamma_e=\dfrac{\nu n m }{\rho s_{tr}}K_0(q,y),
$$
где
$$
K_0(q,y)=1+\dfrac{3y}{2q^3}\varphi_0(q,y).
$$

As it was already specified, coefficient $K_0 (q, y) $, obtained without elect\-ri\-cal
field, the attenuation dimensionless coefficient $K (q, y) $ effectively app\-ro\-xi\-mates.
However, the analysis carried out above shows, that for the thin analysis of a damping
coefficient its representation in the form of the total $K(q,y)=K_0(q, y)+K_1(q,y)$
is required.

The present work represents the kinetic approach to damping coefficient
investigation sound wave into degenerate plasma, somewhat alternative to the approach
Pippard' \cite{Pippard}. Investigation of coefficient is carried out
of attenuations of a soundc wave on the basis of the kinetic (dynamic) approach
of interraction of degenerate electronic gas with oscillations of lattice in given work.

%\newpage

 \bigskip

%\newpage

\begin{center}
  \bf Appendix: calculation of integral
\end{center}

$$
J=\int \dfrac{v_y^2\delta(\varepsilon_F-\varepsilon)}{\nu-i\omega+ikv_x}d^3v.
$$

We calculate the integral $J$. WEe note that
$$
\delta(\varepsilon_F-\varepsilon)=\delta\Big(\dfrac{mv_F^2}{2}-\dfrac{mv^2}{2}\Big)=
\dfrac{2}{m}\delta(v_F^2-v^2)=\dfrac{1}{mv_F}\delta(v_F-v).
$$

Firther we have
$$
J=\int \dfrac{v_y^2\delta(\varepsilon_F-\varepsilon)}{\nu-i\omega+ikv_x}d^3v=
\dfrac{1}{mv_F}\int \dfrac{v_y^2\delta(v_F-v)d^3v}{\nu-i\omega +ikv_x}.
$$

We will use the spherical coordinats
$$
v_x=v\cos \theta,\qquad v_y=v\sin \theta \cos\xi,\qquad v_z=v\sin \theta\sin \xi,$$$$
d^3v=v^2 d\mu d\theta d \xi, \qquad \mu=\cos \theta.
$$
Therefore
$$
J=\dfrac{1}{mv_F}\int\limits_{-1}^{1}(1-\mu^2)d\mu \int\limits_{0}^{2\pi} \cos^2\xi
d\xi \int\limits_{0}^{\infty}\dfrac{v^4\delta(v_F-v)dv}{\nu-i\omega+ik v \mu}=
$$
$$
=\dfrac{\pi v_F^3}{m}\int\limits_{-1}^{1}\dfrac{(1-\mu^2)d\mu}{\nu-i\omega+ik v_F
\mu}=-i\dfrac{\pi v_F^2}{mk}\int\limits_{-1}^{1}\dfrac{(1-\mu^2)d\mu}{\mu-
\dfrac{\omega+i \nu}{kv_F}}=
$$
$$
=-i\dfrac{\pi v_F^2}{m k}\int\limits_{-1}^{1}\dfrac{(1-\mu^2)d\mu}{\mu-z/q}.
\eqno{(A.1)}
$$
Here
$$
z=\Omega+iy=\dfrac{\omega+i \nu}{k_Fv_F}, \quad q=\dfrac{k}{k_F}, \quad
\Omega=\dfrac{\omega}{k_Fv_F},\quad y=\dfrac{\nu}{k_Fv_F}.
$$
It easy see that
$$
\int\limits_{-1}^{1}\dfrac{(1-\mu^2)d\mu}{\mu-z/q}=
-2\dfrac{z}{q}+\Big[1-\Big(\dfrac{z}{q}\Big)^2\Big]\ln\dfrac{1-z/q}{-1-z/q}=
$$
$$
=-\dfrac{1}{q^2}\left\{2zq-[q^2-z^2]\ln\dfrac{z-q}{z+q}\right\}=
$$
$$
=-\dfrac{1}{q^2}\left\{2q(\Omega+iy)-
[q^2-(\Omega+iy)^2]\ln\dfrac{\Omega+iy-q}{\Omega+iy+q}\right\}.
$$

We denote
$$
J_0=2q(\Omega+iy)+[q^2-(\Omega+iy)^2]\ln\dfrac{\Omega+iy-q}{\Omega+iy+q}.
$$

Then
$$
\int\limits_{-1}^{1}\dfrac{(1-\mu^2)d\mu}{\mu-z/q}=-\dfrac{J_0}{q^2}.
\eqno{(A.2)}
$$

Substituting (A.1) in (A.2), we gain, that the integral $J $ is equal
$$
J=-i\dfrac{\pi v_F^2}{mk}\cdot \Big(-\dfrac{J_0}{q^2}\Big)
=\dfrac{i\pi v_F^2}{mk_Fq^3}J_0,
\eqno{(A.3)}
$$
or
$$
J=\dfrac{i\pi v_F^2}{mk_Fq^3}\left\{2q(\Omega+iy)-
[q^2-(\Omega+iy)^2]\ln\dfrac{\Omega+iy-q}{\Omega+iy+q}\right\}.
\eqno{(A.4)}
$$

We consider the expression
$$
J_0=2q(\Omega+iy)-[q^2-(\Omega+iy)^2]\ln\dfrac{\Omega+iy-q}{\Omega+iy+q}.
\eqno{(A.5)}
$$

We denote
$$
Z=\dfrac{\Omega+iy-q}{\Omega+iy+q},\quad \ln Z=\ln|Z|+i\arg Z.
$$

It easy see that
$$
\ln\dfrac{\Omega+iy-q}{\Omega+iy+q}=\ln\dfrac{y+i(q-\Omega)}{y-i(q+\Omega)}=$$$$=
\ln(y+i(q-\Omega))-\ln(y-i(q+\Omega))=
$$
$$
=\dfrac{1}{2}\ln\dfrac{(q-\Omega)^2+y^2}{(q+\Omega)^2+y^2}+
i\Big(\arctg\dfrac{q-\Omega}{y}-\arctg\dfrac{-(q+\Omega)}{y}\Big)=
$$
$$
=W_0+iW_1,
$$
where
$$
W_0=\ln|Z|=\dfrac{1}{2}\ln\dfrac{(\Omega-q)^2+y^2}{(\Omega+q)^2+y^2},
$$
$$
W_1=\arg Z=\Big(\arctg\dfrac{q-\Omega}{y}+\arctg\dfrac{q+\Omega}{y}\Big).
$$

We separate in  (2.13) real and imafginary parts
$$
J_0=\Re J_0+i\Im J_0=J_0'+iJ_0'',
$$
where
$$
J_0'=\Re J_0=2q\Omega-(q^2-\Omega^2+y^2)W_0-2y\Omega W_1,
$$
$$
J_0''=\Im J_0=2qy-(q^2-\Omega^2+y^2)W_1+2y\Omega W_0.
$$

Finally, the real and imaginary parts of integral $J$ are equal
$$
\Re J=-\dfrac{\pi v_F^2}{mk_Fq^3}\Bigg[2qy-(q^2-\Omega^2+y^2)\Big(\arctg
\dfrac{q-\Omega}{y}+\arctg\dfrac{q+\Omega}{y}\Big)+$$$$+\dfrac{y}{2}\ln
\dfrac{(\Omega-q)^2+y^2}{(\Omega+q)^2+y^2}\Bigg]
$$
and
$$
\Im J=\dfrac{\pi v_F^2}{mk_Fq^3}\Bigg[2q\Omega-(q^2-\Omega^2+y^2)
\dfrac{1}{2}\ln\dfrac{(\Omega-q)^2+y^2}{(\Omega+q)^2+y^2}
-$$$$-2y\Big(\arctg\dfrac{q-\Omega}{y}+\arctg\dfrac{q+\Omega}{y}\Big)\Bigg].
$$
%\clearpage

\end{document}